\documentclass[12pt]{iopart}
\usepackage{graphicx}
\usepackage{epsfig}
\usepackage{here}
\usepackage{dcolumn}
\usepackage{bm}
\begin{document}
\title{Exact Floquet states of a driven condensate and their
stabilities}

\author{Wenhua Hai$^1$, Chaohong Lee$^2$ and Qianquan Zhu$^1$}

\address{$^1$Department of Physics, Hunan Normal University, Changsha 410081, P. R. China\\
$^2$Nonlinear Physics Centre and ARC Centre of Excellence for
Quantum-Atom Optics, Research School of Physical Sciences and
Engineering, Australian National University, Canberra ACT 0200,
Australia}

\eads{\mailto{ whhai2005@yahoo.com.cn (corresponding author)},
\mailto{chl124@rsphysse.anu.edu.au}}

\begin{abstract}
We investigate the Gross-Pitaevskii equation which describes an
atomic Bose-Einstein condensate confined in an optical lattice and
driven by a spatiotemporal periodic laser field. It is demonstrated
that the exact Floquet states appear when the external
time-dependent potential is balanced by the nonlinear mean-field
interaction. The balance region of parameters is divided into a
phase-continuing region and a phase-jumping one. In the latter
region, the Floquet states are spatiotemporal vortices of nontrivial
phase structures and zero-density cores. Due to the velocity
singularities of vortex cores and the blowing-up of perturbed
solutions, the spatiotemporal vortices are unstable periodic states
embedded in chaos. The stability and instability of these Floquet
states are numerically explored by the time evolution of fidelity
between the exact and numerical solutions. It is numerically
illustrated that the stable Floquet states in the phase-continuing
region could be prepared from the uniformly initial states by slow
growth of the external potential.
\end{abstract}



\submitto{\jpb}

\maketitle

\section{Introduction}

The atomic Bose-Einstein condensates (BECs) in optical lattices
have stimulated great interests in both the many-body quantum
effects \cite{Folling}-\cite{Lee3} and mean-field dynamics
\cite{WLiu}-\cite{Elena2}. It has been suggested for applications
in quantum interference \cite{Folling,Anderson}, quantum
information processing \cite{Pachos,Ionicioiu,Lee2} and
matter-wave manipulation \cite{Ovchinnikov,Choi}. Using the
mean-field theory, some exact stationary solutions for the systems
of one-dimensional (1D) optical lattices were obtained
\cite{Bronski, Deconinck1,hai}, and their stabilities are
discussed \cite{Wu,Bronski2}. Some nonstationary states of BECs in
time-dependent lattices have also been investigated
\cite{Ruprecht}-\cite{Deng}.

The Floquet states, a kind of nonstationary states, have been
extensively introduced to understand the dynamics of various
driven systems \cite{Holthaus,QXie}. For a linear Schr\"{o}dinger
system of $T$-periodic Hamiltonian, $H(\vec{r},t) =
H(\vec{r},t+nT)$, the Floquet theorem allows one to write its
states as $\Psi(\vec{r},t) = U(\vec{r},t) \exp (-iE_F t)$ with
periodic function $U(\vec{r},t) = U(\vec{r}, t+nT)$ and
quasienergy $E_F$. The Floquet analysis is analogous to the Bloch
analysis in solid state physics in which the states of spatially
periodic system are written in terms of Bloch states and
quasimomenta. Recently, the Floquet analysis has been introduced
to systems of condensed atoms. For driven systems of BECs in
double-well potentials, applying a two-mode approximation to the
Gross-Pitaevskii (GP) equation, the Floquet states have been used
to analyze the coherent control of the population self-trapping
\cite{Holthaus,QXie}. For the driven ultracold Bose atoms in
optical lattices, which obey driven Bose-Hubbard models in full
quantum theory, the Floquet states are applied to investigate the
dynamical superfluid-insulator transition \cite{Eckardt}. Below,
we will generalize the familiar Floquet states in driven linear
Schr\"{o}dinger systems to the ones in a driven nonlinear
Schr\"{o}dinger system without any approximation.

Spatial vortices \cite{pismen} are fundamental objects of
spinning, often turbulent, flow (or any spiral motion) with closed
streamlines. They widely exist in different fields including phase
singularities in optics \cite{soskin} and circulating particles in
superfluids and BECs \cite{bec}. Recently, the conception of
vortex in spatial domain has been generalized to the one in
spatiotemporal domain. It shows that the superposition of two
phase modulated optical beams generates the train of
spatiotemporal vortices, which are periodical in space or time
\cite{SPV}. We expect such spatiotemporal vortices can also appear
in an atomic BEC under some certain conditions.

The stability analysis can provide useful information for
preparation, control and application of a particular state.
Mathematically, the instability appears when an initially small
deviations do not keep bounded. It has been demonstrated that a
blowing-up solution \cite{Guo} appears in a BEC system and is
related to the BEC collapse and instability \cite{Konotop}. For a
linear quantum system, the fidelity between the unperturbed and
perturbed states, a quantum Loschmidt echo, has been successfully
used to analyze the stability \cite{Peres,Gorin}. In which, the
fast decay of fidelity corresponds to the instability of a quantum
evolution. In the following, we will extend such a fidelity
analysis to the nonlinear quantum systems of BECs.

In this article, we show how to prepare the exact Floquet states of
an atomic BEC trapped in an optical lattice and driven by a
spatiotemporally periodic field, and also analyze their stabilities.
Utilizing the balance condition between the external potentials and
mean-field interaction, we obtain a kind of exact Floquet solutions
for the non-integrable chaotic system. The balance region of
parameters is divided into the phase-continuing region and
phase-jumping one, which are associated with the stable periodic
states and unstable spatiotemporal vortex states embedded in chaos
respectively. The stability and instability are confirmed by the
corresponding fidelities between the exact and the numerical
solutions. Our results suggest a method for suppressing the
instability and preparing stable non-stationary states of the
condensates. The considered condensate stabilization and preparation
could be experimentally realizable.

\section{Model and exact Floquet states}

We consider an atomic BEC with strongly transverse confinement,
and so that it obeys a quasi-1D GP equation,
\begin{eqnarray}\label{1}
i\hbar\psi_t = -\frac{\hbar^{2}}{2m}\psi_{xx} + [g_{1d}|\psi|^2 +
V (x,t)] \psi,
\end{eqnarray}
where $m$ is the single-atom mass and $g_{1d}$ denotes the
quasi-1D interaction strength \cite{Abdullaev}, and $V(x,t)$
stands for the external potential $V(x,t) = V_0\cos^2 (kx) +
f(x,t)$ with lattice potential $V_0\cos^2 (kx)$ of strength $V_0$
and driving field $f(x,t)$ to be determined, and the wave vector
$k$.

To obtain the exact solutions of model (1) with the balance
technique \cite{Bronski,hai}, we use the balance condition
\begin{equation}
g_{1d}|\psi|^2 + V (x,t) = E_F
\end{equation}
with $E_F$ being a constant, which means that the nonlinear
mean-field interaction is balanced by the external potential.
Subsequently, the model (1) is simplified as
\begin{equation}
i\hbar\psi_t=-\frac{\hbar^{2}}{2m}\psi_{xx}+E_F\psi
\end{equation}
whose complete solution which contains some arbitrary constants can
be easily obtained \cite{hai} and the corresponding integration
constants are determined by inserting the complete solution into Eq.
(2). Obviously, Eqs. (2) and (3) supports the exact Floquet state,
\begin{eqnarray}
\psi(x,t)&=&\left [\sqrt{\frac{E_F}{g_{1d}}}+\alpha
\sqrt{-\frac{V_0}{g_{1d}}}\cos
(kx) e^{-i\omega t}\right ] \nonumber \\
&&\times \exp \left(-\frac{i}{\hbar} E_F t \right ),
\end{eqnarray}
where $\omega$ is the driving frequency, $E_F$ is the Floquet
energy, $\sqrt{E_F/g_{1d}}$, $\sqrt{-V_0/g_{1d}}$ and
$\sqrt{-E_FV_0}$ are real quantities, and $\alpha$ denotes a
constant with value being either $1$ or $-$1. In the absence of a
potential, $V (x,t) =0$, the solution has an uniform density
distribution $|\psi|^2 = \mu/g_{1d}$ for $\psi=\sqrt{\mu/g_{1d}}\
e^{-i\mu t}$ with $\mu=E_F$. In the case of $V_0 \ne 0$ and $f=0$,
the balance solution (4) contains the stationary state
$\psi(x,t)=\alpha \sqrt{-V_0/g_{1d}}\cos (kx) e^{-i\mu t}$ for the
chemical potential $\mu=\omega=\hbar k^2/2m$ and zero Floquet energy
$E_F=0$. In general, the balance condition (2) and Floquet solution
(4) require the driving frequency $\omega$ and recoil energy $E_r$
of the optical lattice satisfy the relation
\begin{equation}
\hbar \omega=E_r=\frac{\hbar^2k^2}{2m},
\end{equation}
and the driving field has the form
\begin{equation}
f(x,t)=V_1\cos kx\cos\omega t,\ \ \ V_1=2\alpha \sqrt{-E_FV_0},
\end{equation}
where the driving strength $V_1$ is determined by the lattice depth
$V_0$ and the Floquet energy $E_F$. Given Eq. (5) and the form of
potential, the second of Eq. (6) becomes the balance condition
corresponding to Eq. (2). This condition indicates the relation
between the driving strength $V_1$ and frequency $\omega$ for the
requirement of the balance. The real values of $\sqrt{E_F/g_{1d}}$
requires that the attractive ($g_{1d}<0$) and repulsive ($g_{1d}>0$)
condensates have negative and positive Floquet energies,
respectively. The driving field in Eq. (6) is a time-dependent laser
standing wave, which can be formed from the linear superposition of
two counter-propagating travelling waves \cite{Fallani}. It is clear
that the classical system governed by the total potential
$V(x,t)=V_0\cos^2 kx+V_1\cos kx\cos\omega t$ is chaotic and
non-integrable \cite{Bishop,Hensinger,Yang}, and the corresponding
GP system (1) is also non-integrable chaotic one even for the
stationary state case \cite{Chong1}.

Starting from the non-stationary states (4), one can selectively
prepare different stationary states of expected chemical potential
$\mu=\omega$ via selecting the driving frequency $\omega$ and
adiabatically switching off the driving field, namely decreasing
$V_1$ to zero very slowly. Under the balance condition (6) this
adiabatic operation makes $V_1=0$ and $E_F=0$ such that Eq. (4)
becomes the simple stationary solution $\psi(x,t)=\alpha
\sqrt{-V_0/g_{1d}}\cos (kx) e^{-i\mu t}$ for $\mu=\omega$ and
$\alpha=\pm 1$, which still obeys the balance condition (2) and
additional Eqs. (3) and (5). It has been demonstrated that such a
stationary state which has zero points possesses instability or
undetermined stability for $g_{1d}>0$ or $g_{1d}<0$ respectively
\cite{Bronski2}. Different from the stationary states, the exact
Floquet state for a driven condensate has a nontrivial phase which
cannot be simply decoupled into a spatially dependent part from the
super flow and a temporally dependent part dominated by the chemical
potential. In contrast, this nontrivial phase includes a
spatiotemporally dependent part from the super flow and a temporally
dependent part dominated by the Floquet energy. This is resulted
from the spatiotemporal dependence of the super flows in driven
condensates.

Defining the average number of atoms per well as
$N=\int_0^{\pi}|\psi|^2d(kx)/k=\pi(E_F-V_0/2)/(kg_{1d})$, the
Floquet energy can be expressed as $E_F=kg_{1d}N/\pi+V_0/2$. The
real values of $\sqrt{E_F/g_{1d}}$ and $\sqrt{-V_0/g_{1d}}$ in Eq.
(4) require $E_F/g_{1d}=kN/\pi+V_0/(2g_{1d})\ge 0$ and the lattice
strength $V_0$ satisfying,
\begin{equation}
|V_0|\le V_c = 2kN|g_{1d}|/\pi.
\end{equation}
Given the $E_F$ and condition $V_0/g_{1d}<0$, the balance
condition in Eq. (6) gives $V_1=2\alpha \sqrt{-E_FV_0}
=\alpha\sqrt{-2(2kNg_{1d}/\pi+V_0)V_0}$. This formula and Eq. (7)
confine the driving field strength
\begin{equation}
|V_1| = \sqrt{2|V_0|(V_c-|V_0|)} \le V_c/\sqrt{2},
\end{equation}
where the inequality is derived from the maximum condition
$d|V_1|/d|V_0|=0$ for a fixed $V_c$. Therefore, $V_c$ and
$V_c/\sqrt{2}$ are the supercritical values of the lattice depth
$|V_0|$ and driving strength $|V_1|$ for the balance solution.

Due to $\cos (kx) = \frac 1 2 (e^{ikx}+e^{-ikx})$, the exact
Floquet solution (4) can be regarded as the coherent superposition
between an atom standing wave formed by two counter-propagating
plane waves and the background $\sqrt{E_F/g_{1d}}\exp(-i E_F
t/\hbar)$. Writing the macroscopic wave function as
$\psi=R(x,t)\exp[i\theta(x,t)]$, the exact solution (4) implies
the atomic-number density
\begin{eqnarray}
R^2&=&|\psi|^2 \\
&=&\frac{1}{g_{1d}}\left ( E_F-V_0\cos^2 kx-V_1\cos kx \cos \omega
t\right )\nonumber
\end{eqnarray}
and the phase
\begin{eqnarray}
\theta&=&\arctan[Im (\psi)/Re(\psi)] \\
&=&\arctan\frac{\sqrt{|V_0|}\cos kx\sin \omega
t}{\sqrt{|E_F|}+\sqrt{|V_0|}\cos kx\cos \omega
t}-\frac{E_F}{\hbar}t.\nonumber
\end{eqnarray}
Clearly, the atomic density $R^{2}(x,t)$ has the same profile of the
potential function and can be controlled by the external fields. The
exact formula for the amplitude $R$ and its finite form will raise
the control precision. The non-integrability of the system means
that its solution cannot contain the integration constants
determined by the initial and boundary conditions. Therefore, the
general solution cannot be given exactly. Of course, one may write
an exact solution in terms of a superposition of an ensemble of
complete orthogonal sets including Fourier modes, which could
contain infinite terms.

\section{Spatiotemporal vortices}

It is well known that the vortices are associated with the phase
singularities. The phase accumulation along a closed circle around
the vortex core and the phase jump along a line through the vortex
core are determined by the vortex charge. According to the
relation between the phase and the velocity,
$\vec{v}=\frac{\hbar}{m}\vec{\nabla} \theta$, the velocity
divergence appears when there is phase jump. In our GP system (1),
the velocity field is defined as $v=\frac{\hbar}{m}\theta_x$ with
\begin{eqnarray}
\theta_x=\frac{kV_1}{2g_{1d}R^2(x,t)}\sin kx\sin \omega t
\end{eqnarray}
denoting the first derivative (gradient) of the phase (10). The
corresponding flow density $J=R^2\theta_x=
\frac{kV_1}{2g_{1d}}\sin kx\sin \omega t$ with amplitude being
proportional to the driving strength $V_1$ describes the Floquet
oscillations of the system. Given the density distribution (9), we
know that the macroscopic wave function has zero-density nodes
$(x_{zd},t_{zd})$ satisfying $R^2(x_{zd},t_{zd}) = 0$, i.e.,
\begin{equation}
\cos (kx_{zd}) = \frac{-V_1\cos \left(\omega t_{zd}\right) \pm
\sqrt{V_1^2\left[\cos^2 \left(\omega
t_{zd}\right)-1\right]}}{2V_0}.
\end{equation}
Because $\cos (kx_{zd})$ are real quantities, one can get $$\cos
(\omega t_{zd}) = \pm 1, ~ \cos (kx_{zd}) = -\frac{V_1}{2V_0}\cos
(\omega t_{zd})=\pm \frac{V_1}{2V_0}$$ from the above formula.
Thus the spatiotemporal coordinates $(x_{zd},t_{zd})$ for the
zero-density nodes read
\begin{eqnarray}
t_{zd} &=& t_n = n \frac{\pi}{\omega}, \\
x_{zd} &=& x_{nl}^{\pm} = \frac 1 k \left[\pm \arccos \left(
\frac{-V_1\cos n\pi}{2V_0} \right)+2l\pi\right]
\end{eqnarray}
with non-negative integers $n$ and $l$. The phase gradient (11)
becomes infinite at those zero-density nodes when $\sin \left(
kx_{nl}^{\pm} \right)\ne 0$. The infinite phase gradient
$\theta_x(x_{nl}^{\pm},t_n)$ means that the phase is a step
function across the zero-density nodes $(x_{nl}^{\pm},t_n)$. Using
the relation between velocity and phase gradient, we know that the
velocity singularity appears at these points. Due to the nonlinear
mean-field interactions otherwhere, Eqs. (11) and (9) imply that
the velocity singularity at the density nodes with
$E_F=V(x_{zd},t_{zd})$ is an effect of nonlinear resonance
\cite{Adhikari}.

The conditions (8) and $\sin \left( kx_{nl}^{\pm} \right) \ne 0$
which is equivalent to $\cos^2 \left( kx_{nl}^{\pm} \right)
=\frac{V_1^2}{4V_0^2}<1$ give the following parametric region,
\begin{eqnarray}
|V_1|<2|V_0|\ \ \ \ and \ \ \ \ V_c/3<|V_0|<V_c,
\end{eqnarray}
in which phase jumps may occur. In this parametric region,
applying the l'H\"{o}pital rule to the phase gradient (11), one
can find
\begin{equation}
\lim_{t\to t_n}\theta_x(x_{nl}^{\pm},t)=\pm \frac{k}{2}\lim_{t\to
t_n}\left[\tan (kx_{nl}^{\pm})\cot (\omega t)\right]=\infty.
\end{equation}
However, the time derivative of phase,
\begin{equation}
\lim_{t \to t_n}\theta_t(x_{nl}^{\pm},t) = \omega/2-E_F/\hbar
\end{equation}
keeps finite at $(x_{nl}^{\pm},t_n)$. This means that the phase
jumps occur along the spatial direction but not along the temporal
direction. Thus, the whole balance region is divided into two
parts with or without phase jumps. The boundary between two parts
has $|V_1|=2\sqrt{|E_FV_0|}= 2|V_0|$, $|V_0|= |E_F|=V_c/3$,
$R^2(x_{nl}^{\pm},t_n)=0$, $\sin (kx_{nl}^{\pm}) =0$ and finite
$\theta_x(x_{nl}^{\pm},t_n)$. The latter indicates that the exact
Floquet solutions at the boundary are not phase-jump solutions. In
the phase-jump region, the driving field strength $|V_1|$ is
smaller than two times of the lattice strength $2|V_0|$.

\begin{figure}[htp]
\center
\includegraphics[width=4.2in]{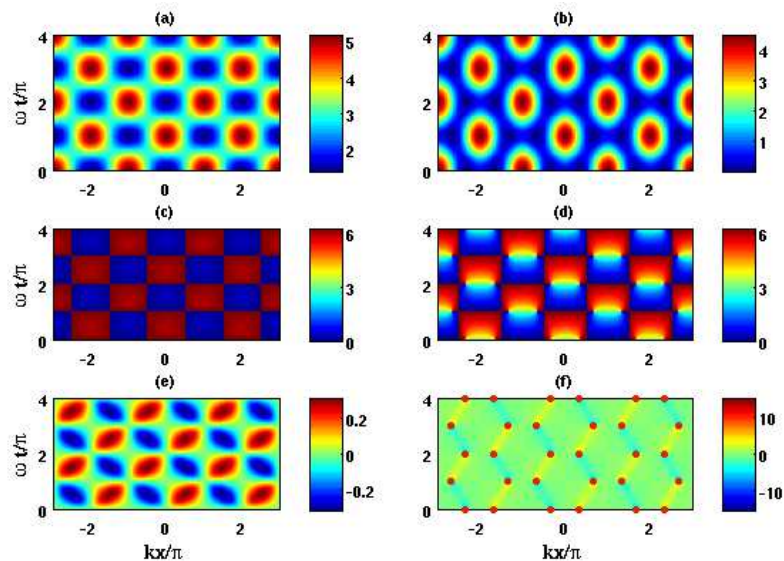}
\caption{The spatiotemporal evolutions of the exact Floquet
states: density evolutions (a) and (b), phase evolutions (c) and
(d), and velocity evolutions (e) and (f). Left column [(a), (c)
and (e)] corresponds to parameters $E_F/g_{1d}=3 k, \
V_0/g_{1d}=-0.3 k, \ V_1/g_{1d}=1.8974 k$ in phase-continuing
region $|V_1|>2|V_0|$. Right column [(b), (d) and (f)] is
associated with $E_F/g_{1d}=0.5 k, \ V_0/g_{1d}=-2 k, \
V_1/g_{1d}=2 k$ in phase-jumping region $|V_1|<2|V_0|$. The red
dots in (f) are the points where phase jumps and velocity
divergence occur. The atomic number-density $R^{2}$ is normalized
in units of $k$, and the phase contribution from the factor $\exp
(-\frac{i}{\hbar}E_F t)$ is eliminated.}\label{fig1}
\end{figure}

Similarly, applying the l'H\"{o}pital rule to the following
quantities, one can easily obtain
\begin{eqnarray}
& &\lim_{t \to t_n} R_x(x_{nl}^{\pm},t)\propto R(x_{nl}^{\pm},t_n)=0,\nonumber\\
& &\lim_{t \to t_n} R_t(x_{nl}^{\pm},t) = V_1\omega_1/\sqrt{V_0},\nonumber \\
& &\lim_{t \to t_n} R_{xx}(x_{nl}^{\pm},t)\propto
R^{-1}(x_{nl}^{\pm},t_n) =
\infty,\nonumber \\
& &\lim_{t \to t_n} \theta_{xx}(x_{nl}^{\pm},t) \propto
R^{-1}(x_{nl}^{\pm},t_n) = \infty.\nonumber
\end{eqnarray}
With these formulas, one has $$\lim_{t\to
t_n}\psi_t(x_{nl}^{\pm},t)\propto R_t(x_{nl}^{\pm},t_n)=
constant,$$ and
$$\lim_{t\to t_n}\psi_x(x_{nl}^{\pm},t)\propto
\theta_x(x_{nl}^{\pm},t_n)R(x_{nl}^{\pm},t_n)=constant.$$ Since
$\psi, \ \psi_t$ and $\psi_x$ are bounded, the phase-jump solution
$\psi$ is a bounded solution rather than a blowing-up one
\cite{Guo}.

In Fig. 1, we show the spatiotemporal evolutions of the exact
Floquet states for different parameters. The first, second and
third rows show density, phase and velocity evolutions,
respectively. The left and right columns have the parameters out
and in region of phase jumps, respectively. For simplicity,
without loss of generality, we rescale the density in units of $k$
and eliminate the phase from the factor $\exp (-\frac{i}{\hbar}E_F
t)$. It clearly shows that the phase-jump solution (right column)
has nodes of zero density at which $\pi$ phase jumps occur along
the spatial direction and the corresponding velocities approach to
infinite.

The nontrivial phase structure around the singular points, that is,
the circulation integral $\oint (\theta_x dx+\theta_t dt)=\oint
d\theta=2n\pi$ ($n=0,1,2\cdots$) along closed spatiotemporal
trajectories enclosing a singular point, reminds us the well-known
Onsager-Feynman quantization condition for planar vortices
\cite{Onsager, Penn}, since Eq. (16) implies that the integral
around the spatiotemporal singular points is not equal to zero, and
Eq. (10) allows the transformation $\theta\to \theta \pm n\pi$ at
any spatiotemporal point. Therefore, mathematically, the circulation
around the singular points turns out to be quantized as a
consequence of the form of field $\theta(x,t)$. This means that the
nonzero quantized integrals along the closed trajectories around the
zero-density nodes indicate the existence of the $(1+1)$-D
spatiotemporal vortex. Due to the analogues between the Floquet and
Bloch analysis, the above Floquet states for an atomic BEC in 1D
optical lattice with a spatiotemporal driving field are similar to
the nonlinear Bloch modes in 2D periodic potentials \cite{Trager},
where the phase-jump solutions for the relatively weak driving
strengthes correspond to the vortex solitons \cite{Alexander}. This
peculiar type of vortices are called as spatiotemporal vortices
\cite{SPV}. Then the zero-density nodes $(x_{nl}^{\pm},t_n)$ are
vortex cores. From the phase distribution in the right column of
Fig. 1, we find the phase accumulation along a circle around the
vortex cores $(x_{nl}^{\pm},t_n)$ is $\pm 2\pi$ . This means that
the vortex charge is $\pm 1$ and so that the vortices at a pair of
$(x_{nl}^{+},t_n)$ and $(x_{nl}^{-},t_n)$ are a vortex-antivortex
pair.

\section{Stability analysis and state preparation}

In this section, we analyze the stability of the exact Floquet
solution (4) for the nonlinear quantum system (1). In the sense of
Lyapunov's stability, the instability entails that the initially
small deviations grow in time without upper limit. Within the
linear stability analysis \cite{Bronski2,Wu}, the deviations are
governed by the linearized equations for the unperturbed solution.
If the linearized equations have any unbounded solution in time
evolution, the corresponding unperturbed state is unstable. Only
if all perturbed solutions are bounded, the unperturbed solution
is stable. Usually, it is difficult to explore the stability of a
nonstationary solution. However, it is relatively easy to show its
instability. We will demonstrate that whether the nonlinear
resonance from the velocity divergence can cause the dynamical
instability.

The perturbations to the exact Floquet solution (4) can be taken
in different forms. We consider the one of them, which has been
used for stability analysis of stationary state solutions
\cite{Bronski2}, in form of
\begin{eqnarray}
\psi=[R(x,t)+\varepsilon \psi_1(x,t)]\exp [i\theta(x,t)-iE_F
t/\hbar],
\end{eqnarray}
where the small parameter $\varepsilon$ obeys $|\varepsilon|\ll 1$
and the perturbation correction $\psi_1(x,t)=\phi(x,t)+i
\varphi(x,t)$ including a real part $\phi$ and an imaginary part
$\varphi$. According to the linear stability analysis, the
Lyapunov stability of solution requires that the unperturbed and
perturbed solutions satisfy the spatially boundary conditions and
all first-order corrections including $||\psi_1||
=\sqrt{|\phi|^2+|\varphi|^2}$ for arbitrarily boundary conditions
are bounded in the time evolution, and the Lyapunov instability is
associated with the unboundedness of any perturbed solution.
Substituting the perturbed solution (18) into the original
equation (1), we get the linearized equations
\begin{eqnarray}
\hbar\phi_t&=&L_1 \varphi-S\phi, \\
\hbar\varphi_t&=&-(L_3\phi+S\varphi),
\end{eqnarray}
where the operators $L_j$ and $S$ satisfy$^{\cite{Bronski2}}$
\begin{eqnarray}
L_j&=&-\frac {\hbar^2}{2m} \Big[\frac{\partial ^2}{\partial x
^2}-\theta_x^2(x,t)\Big]+jg_{1d}R^2(x,t)\nonumber \\
&&+V(x,t)+\hbar\theta_t(x,t), (for \ j=1 \ and \ 3),\\
S&=&\frac {\hbar^2}{2m}\Big[2\theta_x(x,t)\frac{\partial}{\partial
x}+ \theta_{xx}(x,t)\Big].
\end{eqnarray}
Using these operators and the relation $\psi=R\exp (i\theta)$, we
can rewrite the unperturbed system (1) as $L_1R=0$ and $\hbar
R_t+SR=0$.

From the linearized equations (19)-(22), we know that the bounded
solutions $\phi$ and $\varphi$ must be zero at the zero-density
nodes $(x_{nl}^{\pm},t_n)$ which are singular points of $\theta_x$
and $\theta_{xx}$. It is quite difficult to solve the perturbed
equations (19) and (20) for $\phi$ and $\varphi$. However, we can
show the instability from the singularity of first derivatives
$\phi_t$ and $\varphi_t$. Mathematically, the linearized equations
(19) and (20) are a couple of partially differential equations with
general solutions containing some arbitrary functions, which are
associated with various initial perturbations. For the solutions of
spatiotemporal vortices, the linearized equations (19) and (20) are
singularly differential equations with infinite $\theta_x$ and
$\theta_{xx}$ at the zero-density nodes $(x_{nl}^{\pm},t_n)$.
Perhaps one may find a set of special solutions for the linearized
equations (19) and (20) whose first time derivatives $\phi_t$ and
$\varphi_t$ are bounded. Nevertheless, since
$\phi(x_{nl}^{\pm},t_n)$ and $\varphi(x_{nl}^{\pm},t_n)$ cannot keep
vanished for the arbitrary functions included in the general
solutions $\phi(x,t)$ and $\varphi(x,t)$, we cannot guarantee the
boundedness of the first time derivatives in the singular Eqs. (19)
and (20). Hence, under some initial perturbations the singularity of
$\theta_x(x_{nl}^{\pm},t_n)$ and $\theta_{xx}(x_{nl}^{\pm},t_n)$
will result in the divergence of $\phi_t(x_{nl}^{\pm},t_n)$ and
$\varphi_t(x_{nl}^{\pm},t_n)$. That is, the linearized equations
(19) and (20) have blowing-up solutions \cite{Guo} in the phase-jump
region which supports trains of spatiotemporal vortex-antivortex
pairs. The blowing-up of the perturbed solutions could break down
the integrability of linearized equations (19) and (20) and induce
the jumps in $\phi(x_{nl}^{\pm},t_n)$ and
$\varphi(x_{nl}^{\pm},t_n)$. Moreover, the jumping heights
$\triangle \phi(x_{nl}^{\pm},t_n)$ and $\triangle
\varphi(x_{nl}^{\pm},t_n)$ may be uncontrollably large, because of
the non-integrability of the linearized equations (19) and (20) and
the singularities of $\theta_x^2(x,t)\phi(x,t)$ and
$\theta_x^2(x,t)\varphi(x,t)$ at the vortex cores
$(x_{nl}^{\pm},t_n)$. Therefore, the blowing-up of perturbed
solutions relates to the instability of the system.

To confirm the above theoretical analysis for instability, we
numerically integrate the GP equation (1) with the well-developed
operator-splitting method. In our numerical simulation, we input
the initial conditions as the initial values of the corresponding
exact Floquet states and choose the same parameters used in Fig. 1
and the periodic boundary conditions $\psi(x_{max},t) =
\psi(-x_{max},t)$ of $x_{max}=4$. To show the difference between
the exact Floquet solution and the numerical one, we calculate the
time evolution of fidelity $F(t)$ between the exact solution
$\psi_{ex}(x,t)$ without perturbations and the numerical one
$\psi_{num}(x,t)$ with initial white noise,
\begin{equation}
F(t) = \frac{1}{N_C (t)}\left|\int
\psi_{num}^{*}(x,t)\psi_{ex}(x,t) dx \right|^2,
\end{equation}
where the normalization function is defined as
\begin{eqnarray}
N_C (t) = \int \psi_{num}^{*}(x,t)\psi_{num}(x,t) &dx& \nonumber\\
      \times \int \psi_{ex}^{*}(x,t)\psi_{ex}(x,t) &dx&,
\end{eqnarray}
which ensures the fidelity satisfying $0 \le F(t) \le 1$. Here,
the functions $\psi^{*}_{ex}(x,t)$ and $\psi^{*}_{num}(x,t)$ stand
for the complex conjugates of functions $\psi_{ex}(x,t)$ and
$\psi_{num}(x,t)$. Due to the GP system (1) conserves the total
number of atoms, we have $N_C (t) = N_C (0)$. The fidelity
describes the similarity between the two solutions, $F=1$ means
the two solutions are the same and $F=0$ means no similarity
between them. So the fast decay of fidelity indicates that the
initially small deviation from the unperturbed state grows fast.
This is a signature for the instability of a quantum evolution
\cite{Peres,Gorin}.

\begin{figure}[htp]
 \center
 \includegraphics[width=3.8in]{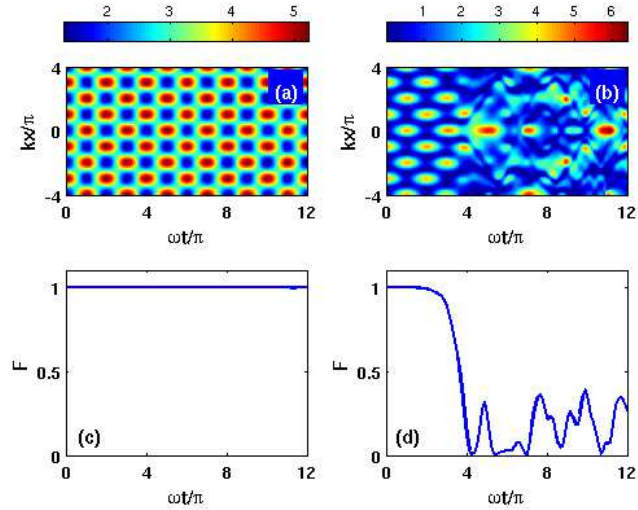}
\caption{The numerical evolutions for the same cases shown in Fig.
1. The left [(a) and (c)] and right [(b) and (d)] columns have the
same parameters with the left and right columns of Fig. 1,
respectively. The above [(a) and (b)] and below [(c) and (d)] rows
show the density and fidelity evolutions, respectively. For the
non-vortex state (left column) the density evolution is regular
and periodic, and the fidelity perfectly keeps close to one. But
for the solution of spatiotemporal vortices (right column) after a
short period of time the density evolution becomes chaotic and the
fidelity shows fast decay, which is a signature of instability.}
 \label{fig2}
\end{figure}

In Fig. 2, we show the numerical simulations correspond to the
exact evolutions shown in Fig. 1. For the non-vortex solutions
(left column), the density evolution from numerical simulation is
regular and periodic, and the fidelity between the numerical and
exact solutions perfectly keeps close to one. This means that the
two solutions are almost the same and the exact Floquet state
maintains its stability under the numerical perturbation. For the
solution of spatiotemporal vortices, the density evolution becomes
chaotic after a short period of time, and the corresponding
fidelity quickly decays to zero after a short period of time and
then chaotically oscillates around a small number close to zero.
This means the spatiotemporal vortices being the periodic state
embedded in chaos and indicates the system losing its stability
under the numerical perturbation. In order to avoid such an
instability, we need adjust the driving field strength $|V_1|$ and
lattice depth $V_0$ to satisfy $|V_1|\ge 2|V_0|$. Due to the
instability is related to the nonlinear resonance, we call such an
instability as the nonlinear resonance instability.

\begin{figure}[htp]
 \center
 \includegraphics[width=3.8in]{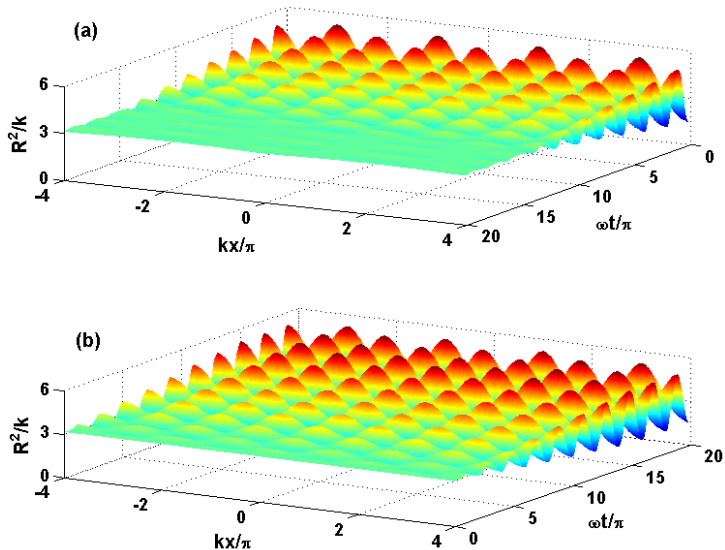}
\caption{State transition and preparation via slow varying
processes. (a) Transition from a stable Floquet state to a uniform
state by slowly ramping down the external potential $V(x,t)= A (t)
[\cos ^2 kx + (V_1/V_0) \cos kx \cos \omega t ]$ from $A=V_0$ to
$A=0$ and then keeping $A$ as zero. (b) Preparation of a stable
Floquet state from an uniform state of density $R^2 (x) =\mu/g_{1d}$
with $\mu=E_F-V_0/2$ and phase $\theta (x) = \theta_0$ by slowly
ramping up the external potential $V(x,t)= A (t) [\cos ^2 kx +
(V_1/V_0) \cos kx \cos \omega t ]$ from $A=0$ to $A=V_0$ and then
keeping $A$ as $V_0$. In which, the values of $V_0$, $V_1$, $E_F$,
$k$ and $\omega$ are as same as the ones for (a) of Fig. 1, and the
linear ramping processes occur in the period from $ t =0$ to $t = 15
\pi/\omega$ and then the amplitude $A$ keeps unchanged after $t = 15
\pi/\omega$.}
 \label{fig3}
\end{figure}

The numerical and analytical results show that the condensate in
the non-stationary states is stable in the parameter region
$|V_1|\ge 2|V_0|, \ V_0\le V_c/3$. Noting the balance condition
$|V_1|=2\sqrt{|E_FV_0|}$ and the average number of atoms per well
$N=\pi(|E_F|+|V_0|/2)/(k|g_{1d}|)$, for a fixed $N$ the Floquet
energy $E_F$ is controlled by the driving field strength $V_{1}$
and lattice depth $V_0$. Therefore, one can control the
oscillation amplitude and flow density of the stable Floquet
states by adjusting the driving field strength and lattice depth.
To precisely prepare the Floquet states analyzed above, one can
use the similar techniques of adiabatic growth which has been
successfully used to prepare stationary states for systems of
time-independent potentials \cite{Bronski}. However, due to our
penitential is time-dependent even ignoring the time-dependence of
the amplitude, the slow varying processes used here is
non-adiabatic. Because of the conserved total particle number, the
state $\psi=R e^{-i\mu t}$ of uniform density distribution $R^2
(x) =\mu/g_{1d}$ with $\mu=E_F-V_0/2$ and trivial phase
distribution $\theta (x) = \theta_0$ ($\theta_0$ is a constant
determined by initial conditions) can be obtained by slowly
ramping down the external potential $V(x,t)= A (t) [\cos ^2 kx +
(V_1/V_0) \cos kx \cos \omega t ]$ from $A=V_0$ to $A=0$.
Inversely, from the uniform state of density $R^2 (x)
=(E_F-V_0/2)/g_{1d}$ and trivial phase, one can get the exact
Floquet state (4) for the stability region $|V_1|\ge 2|V_0|, \
V_0\le V_c/3$ by slowly ramping up the external potential $V(x,t)=
A (t) [\cos ^2 kx + (V_1/V_0) \cos kx \cos \omega t ]$ from $A=0$
to $A=V_0$.

In Fig. 3, we show our numerical simulation for the state
transition and preparation via slow varying processes. For
convenience, we choose the same values of $V_0$, $V_1$, $E_F$, $k$
and $\omega$ for (a) of Fig. 1, which describes a stable Floquet
state. In panel (a), whose external potential is linearly ramped
down from $A=V_0$ to $A=0$, we show the density evolution of a
condensate evolving from the stable Floquet state [which is shown
in panel (a) of Fig. 1] to an uniform state. Accompanying with the
decrease of $|A|$, the density difference between different
positions becomes smaller and smaller. After $t =15 \pi/\omega$,
there is no external potential, the density $R^2 /k = (E_F
-V_0/2)/(k g_{1d}) \pm 0.1 = 3.15 \pm 0.1$. This means that the
relative density difference $\Delta (R^2)/\langle R^2 \rangle$ is
about $3\%$. The fidelity between the final state and the
corresponding exact uniform state is above $98\%$. In panel (b),
whose external potential is linearly ramped up from $A=0$ to
$A=V_0$, we show the density evolution of a condensate evolving
from the state of uniform density $R^2 (x) =(E_F-V_0/2)/g_{1d}$
and trivial phase to a stable Floquet state. Accompanying with the
growth of $|A|$, the density oscillation becomes more and more
close to the one of the exact Floquet state. After $t =15
\pi/\omega$, the fidelity between the prepared state and the exact
Floquet state for the corresponding parameters is also above
$98\%$. The numerical results may supply a useful benchmark for
preparing and manipulating the stable Floquet states in
experiments.

\section{Conclusions and discussions}

In conclusion, for an atomic BEC in a one-dimensional optical
lattice, we have shown how to prepare the exactly nonlinear Floquet
states in a balance parametric region via a spatiotemporally
periodic driving field. The exact Floquet solution describes the
interference between an atomic standing wave and an uniform
background. The balance condition requires that the sum of internal
and external potentials is equal to the Floquet energy which is
proportional to the atomic number per well and the depth of the
lattice potential. It is shown that the phase of the exact Floquet
solution may be continuous or piecewise continuous, depending on the
ratio between the driving intensity and lattice depth. The
phase-jumping solutions with piecewise continuous phase include the
trains of spatiotemporal vortex-anitvortex pairs which are embedded
in chaos. Applying the linear stability analysis, we have analyzed
the stability and instability of the exact Floquet states. The
instability is related to the blowing-up of the perturbed solutions
and the nonlinear resonance. The stable periodic states and unstable
phase-jumping states have been numerically illustrated by the
corresponding fidelities between the analytically unperturbed
solution and the numerically perturbed solution. Dividing the
balance region of parameters into the stability and instability
subregions, we can selectively prepare the stable periodic states
and unstable spatiotemporal vortex states by adjusting the driving
strength and lattice depth to fit the corresponding subregions, and
prepare the different stationary states by adiabatically switching
off the driving field.

We expect the above exact Floquet states in nonlinear quantum
systems would stimulate experimental interests in investigating and
stabilizing the non-stationary and stationary condensates. It would
be helpful to explore the chaotic dynamics in such non-integrable
systems of spatiotemporal periodic potentials like $V(x,t)$
\cite{Hensinger,Bishop}. For the non-integrable system of driven
condensates, the exact Floquet state is completely determined by the
parameters of external field. Therefore, we can prepare and control
the spatiotemporal structures described by the Floquet solutions via
choosing and adjusting parameters in the balance region. For a small
ratio between the driving intensity and lattice depth one could
explore the instability from velocity singularity via observing the
breakdown of the periodic structure in density distribution
\cite{Strekalov,Fallani}. Utilizing the slow varying processes, one
can selectively prepare the stable Floquet states. Particularly, the
results points out a general route to experimentally stabilizing and
preparing the non-stationary states of the condensate.

Additionally, with the similarity between the Floquet and Bloch
analysis, the well-developed techniques for manipulating the
atomic condensates with laser and magnetic fields offer an
opportunity to study the analogue and difference between Floquet
states and Bloch modes including spatial and spatiotemporal
vortices.

\textbf{Acknowledgment} -- The authors thank Prof. Yuri S. Kivshar
for stimulating discussions and valuable suggestions. This work was
supported by the National Natural Science Foundation of China under
Grant No. 10575034, and the Australian Research Council (ARC).

{}

\end{document}